\documentclass[aps,prd,nofootinbib,floatfix,preprintnumbers,amsmath,amssymb]{revtex4}
\usepackage{graphicx}
\usepackage{multirow}
\usepackage[english]{babel}
\usepackage{amsmath}
\usepackage{graphicx}
\usepackage{color}
\usepackage{amssymb}
\usepackage{hyperref}
\usepackage{dcolumn}
\usepackage{bm}

\begin{document}
\preprint{DCP-13-02}

\title{Higgs decay into two photons from a 3HDM with flavor symmetry.}

\author{
Alfredo Aranda,$^{1,2}$\footnote{Electronic address: fefo@ucol.mx}
Cesar Bonilla,$^{3}$\footnote{Electronic address: rasec.cmbd@gmail.com} 
Francisco de Anda,$^{4}$\footnote{Electronic address: franciscojosedea@gmail.com }
Antonio Delgado,$^{5}$\footnote{Electronic address: antonio.delgado@nd.edu }  
J. Hern\'andez--S\'anchez,$^{2,6}$\footnote{Electronic address: jaimeh@ece.buap.mx}
\vspace*{0.3cm}}

\affiliation{$^1$Facultad de Ciencias, CUICBAS,
Universidad de Colima, Colima, M\'exico \\
$^2$Dual C-P Institute of High Energy Physics, M\'exico \\
$^3$ Facultad de Ciencias F\'\i sico-Matem\'aticas, Benem\'erita Universidad Aut\'onoma de Puebla, M\'exico \\
$^4$ Departamento de Fisica, CUCEI, Universidad de Guadalajara, M\'exico \\
$^5$ Department of Physics, University of Notre Dame, Notre Dame IN 46556, USA \\
$^6$ Fac. de Cs. de la Electr\'onica, Benem\'erita Universidad Aut\'onoma de Puebla, Apdo. Postal 542, 72570 Puebla, Puebla, M\'exico }

\date{\today}

\begin{abstract}
In this short letter we show that the excess of events in the decay of Higgs to two photons reported by ATLAS and CMS can be easily accommodated in a flavor renormalizable three Higgs doublet model (3HDM). The model is consistent with all fermion masses, mixing angles, and flavor changing neutral current constraints.
\end{abstract}
\maketitle

Recent results by ATLAS and CMS suggest an excess in the decay $H\to \gamma\gamma$ with respect to the Standard Model (SM) expectation~\cite{Chatrchyan:2012twa,ATLAS:2012ad}. Since the coupling of the Higgs to photons is radiative one possible reason for the excess, without changing the rest of the decays, is the existence of new charged particles with masses of hundreds of GeV. On the other hand there does not seem to be any significant deviation in the gluon fusion rate nor in the decay rates of the Higgs into gauge bosons, therefore the new particles responsible for the excess on the rate to photons have to be colorless and preferably not spin 1, in order to avoid any mixing with the W or the Z. Thus it seems that the existence of one or more spin 0 or spin 1/2 charged particles could explain the diphoton rate measured at the LHC.

In this short letter we show that it is possible to accommodate the excess in the diphoton rate within a renormalizable three Higgs doublet model (3HDM) whose motivation is to describe, in a minimal way, the flavor structure of the SM. Having extra Higgs doublets imply, upon electroweak symmetry breaking, the existence of new charged spin zero particles, two in this particular model. We are going to show that it is possible to satisfy all flavor constrains in this model as well as to generate an extra contribution to the decay rate of the Higgs to photons due to the two new charged degrees of freedom.


The model: We explore a renormalizable flavor model based on the cyclic group $Z_5$ that contains three Higgs doublets. We chose $Z_5$ as it has been shown to be the smallest Abelian symmetry that in the context of 3HDM leads to the Nearest Neighbour  Interaction (NNI) Yukawa textures in the quark sector~\cite{Aranda:2012bv}. Furthermore, the lepton sector of the model completely reproduces the one described in~\cite{Aranda:2011rt}, where no right-handed neutrinos are introduced and neutrino Majorana masses are generated radiatively through the presence of an SU(2) singlet field charged under both Hypercharge and Lepton number.

The model particle content consists of the Standard Model fermion fields, three SU(2) doublets (Hypercharge $y=1/2$) called ${\cal H} = (H,\Phi_1,\Phi_2)$, and a singlet scalar field $\eta$ with Hypercharge $y=-1$ and Lepton number $L=2$. The $Z_5$ charge assignments for the fermion fields are given by
\begin{eqnarray}
\label{Z5fermioncharges}
\overline{Q}_L \simeq \overline{L}_L \simeq (0,-3,-1), \ \ u_R\simeq d_R\simeq e_R \simeq (3,0,2), 
\end{eqnarray}
where $Q_L$ and $L_L$ denote the quark and lepton left-handed SU(2) doublet SM fields respectively, while $u_R$, $d_R$, and $e_R$ denote the quark and charged lepton right-handed SU(2) singlet SM fields. The scalar sector assignments are given by
\begin{eqnarray}
\label{Z5scalarcharges}
{\cal H} \equiv (H,\Phi_1,\Phi_2) \simeq (0,-1,1), \ \ \eta \simeq (-1).
\end{eqnarray}

These assignments lead to NNI textures for the quarks and charged lepton Yukawa matrices, while the neutrino mass matrix is of the form
\begin{eqnarray}
\label{numassmatrix}
M_\nu= \left(\begin{array}{ccc}
A&B&C\\
B&0&0\\
C&0&D\\
\end{array}\right).
\end{eqnarray}

The scalar potential for ${\cal H}$ is given by (see Eq.~(12) of~\cite{Aranda:2012bv})
\begin{eqnarray}
\label{potential} \nonumber
V(H,\Phi_a) &=& \mu_0^2 |H|^2 + \mu_a^2|\Phi_a|^2 + \mu_{0a}^2\left( \Phi_a^{\dagger}H + h.c \right)
+\mu_{12}^2 \left(\Phi_1^{\dagger}\Phi_2 + h.c \right)
+ \lambda_0 \left( |H|^2\right)^2 \nonumber \\ 
&+&\lambda_a \left( |\Phi_a|^2 \right)^2 
+ \lambda_{0a} |H|^2|\Phi_a|^2 + \lambda_{12} |\Phi_1|^2|\Phi_2|^2 +
\tilde{\lambda}_{ab}|\Phi^{\dagger}_a \tilde{\Phi}_b|^2 +\lambda'_{0a}\Phi_{a}^{\dagger}HH^{\dagger}\Phi_{a}\notag\\
&+&\lambda_{3}\left(\Phi_{1}^{\dagger}H\Phi_{2}^{\dagger}H + h.c\right),
\end{eqnarray}
where $a=1,2$ and the terms proportional to $\mu_{0a}$ and $\mu_{12}$ are $Z_5$ soft breaking terms 
required in order to obtain the correct electromagnetic invariant vacuum. The scalar $\eta$ in the model plays its role in the neutrino sector where it provides the necessary Lepton number violation in order to generate Majorana masses. Due to its heavy mass and small mixing, it does not play a role in the phenomenology of the charged scalars present in ${\cal H}$. When the diagonalization of the scalar sector is performed we thus only consider the potential in Eq.~\eqref{potential}.

As discussed in~\cite{Aranda:2012bv,Aranda:2011rt} the textures obtained in this model can reproduce all the observed masses and mixing angles in the quark and charged lepton number sectors, as well as the squared mass differences (for inverted hierarchy only)  and mixing angles in the neutrino sector. Furthermore, it was also shown in~\cite{Aranda:2012bv} that the model is able to satisfy the strong flavor changing neutral current constraints coming from $K-\overline{K}$ mixing.


Results: A numerical scan of the parameter space of the model was performed in order to find regions where the model successfully reproduces/satisfies i) all quark masses and mixing angles, ii) charged lepton masses, squared neutrino mass differences and lepton mixing angles, iii) EW~\cite{STU} and FCNC bounds from $\overline{K}-K$ mixing. We also impose the lightest CP-even neutral scalar to have a mass of $125$ GeV, this scalar has SM-like copings to gauge bosons but can have reduced couplings to the fermions.

The regions that satisfy all these constraints are then used to determine the contribution to the decay of the lightest neutral CP-even scalar, $h^0$ into two photons. This decay is computed using the expression~\cite{HHG}
\begin{eqnarray}
\label{decayamplitude}
|M|^2 = \frac{g^2m_H^2}{32\pi^2 m_W^2} \bigg| \sum_i \alpha N_c e_i^2F_i\bigg|^2 \ ,
\end{eqnarray}
where $i$ runs over scalars, fermions and vector bosons in the loop with charge $e_i$ and color factor $N_c$, and where the $F_i$ factors are given by
\begin{eqnarray}
\label{ffactors}
F_0 &=& \left[ \tau(1-f(\tau))\right]  G_{abc} \\
F_{1/2} &=& \left[ -2\tau(1+(1-\tau)f(\tau)) \right] G_{abc} \\
F_1 &=& \left[2+3\tau+3\tau(2-\tau)f(\tau)\right] G_{abc} \ ,
\end{eqnarray}
with $\tau \equiv (2m_i/m_H)^2$, 
\begin{eqnarray}
\label{ftau}
f(\tau) = \Bigg\lbrace \begin{array}{lc}
 \left[ \sin^{-1}(\sqrt{(1/\tau)} \right]^2 & {\rm if} \ \tau \geq 1 \ \ \\
-\frac{1}{4}\left[\log\left(\frac{1+\sqrt{1-\tau}}{1-\sqrt{1-\tau}}\right)-i\pi\right]^2 & {\rm if} \ \tau < 1 \ ,
\end{array}
\end{eqnarray}
and where the factors $G_{abc}$ contain the model dependent information on mixing angles and couplings among fields $a$, $b$ and $c$ in the diagram vertices ($G_{abc} = 1 \ \forall \ a,b,c$ for the SM).

Following~\cite{Akeroyd:2012ms} we define the ratio
\begin{eqnarray}
\label{Rgammagamma}
R_{\gamma\gamma} \equiv \frac{{\rm BR}(h^0 \to \gamma\gamma)^{\rm MODEL}}{{\rm BR}(h^0 \to \gamma\gamma)^{\rm SM}}
\end{eqnarray}
and take the best value for $R_{\gamma\gamma}$ obtained by combining in average the measurements presented in~\cite{Chatrchyan:2012twa,ATLAS:2012ad,CMS:2012mua,Aad:2012yq}
\begin{eqnarray}
\label{Rbestvalue}
R_{\gamma\gamma} = 2.1 \pm 0.5 \ .
\end{eqnarray}
The most recent results from ATLAS~\citep{ATLAS-new} quote a somewhat smaller ratio given by $R_{\gamma\gamma} = 1.8 \pm 0.3$ with $m_h= 126.6$~GeV, whereas CMS has not presented any update in this channel.

The possible new contribution to the di-photon decay  depends on the mass of the lightest charged Higgs boson $M_{H_1^+}$. We note that the model is able to reproduce acceptable values of $R_{\gamma\gamma}$ in the range $200$~GeV $\leqslant M_{H_1^+} \leqslant 550$~GeV. As expected, the lower part of the range indicates that the contribution from light charged scalars can enhance the di-photon decay rate. Furthermore, note that the lower bound is larger then the top quark mass $m_{top}$ rendering this scenario safe from the bounds obtained in searches for $t \to H^+ b$, $H^+ \to \tau \nu$, and $H^+ \to jj$.

Another important situation is that even in cases when the charged Higgses are too heavy to have a significant contribution to the diphoton rate there could be an increase in $R_{\gamma\gamma}$ due to the reduced couplings to bottoms and not due to the contribution of the heavy charged scalar.

\begin{figure}[ht!]
\includegraphics[scale=0.8]{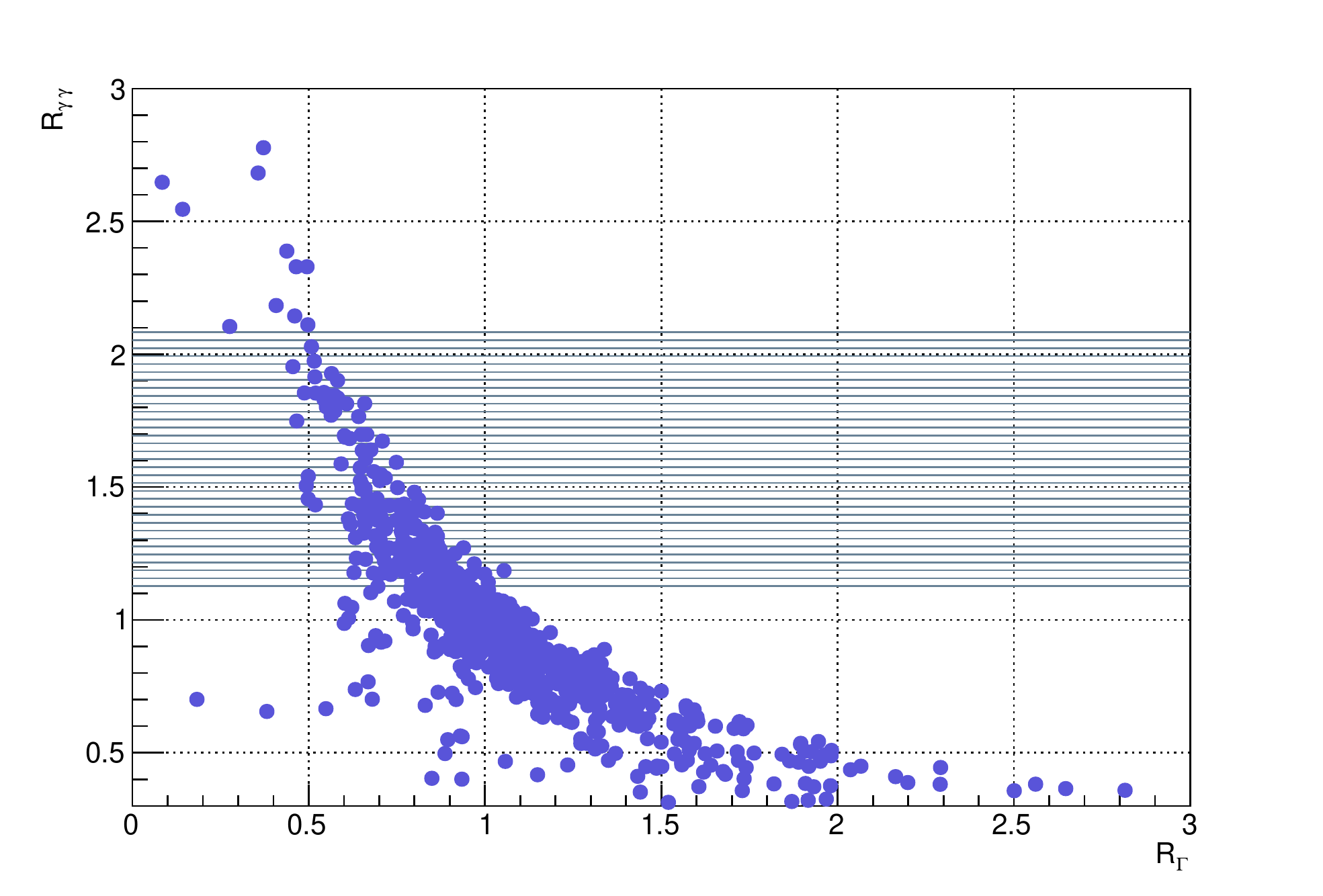}
\caption{Contribution to $R_{\gamma\gamma}$ as a function of the relative width of the lightest Higgs ($R_\Gamma=\Gamma/\Gamma_{SM}$).All points in the plot correspond to cases that reproduce all fermion masses and mixing angles as well as FCNC bounds from $\overline{K}-K$ mixing. The shaded regions correspond to the $1$ $\sigma$ range of $R_{\gamma\gamma}$ as given by~\eqref{Rbestvalue} (gray horizontal lines) and the latest ATLAS results (green vertical lines).}
\label{fig:fig1}
\end{figure}

\begin{figure}[ht!]
\includegraphics[scale=0.8]{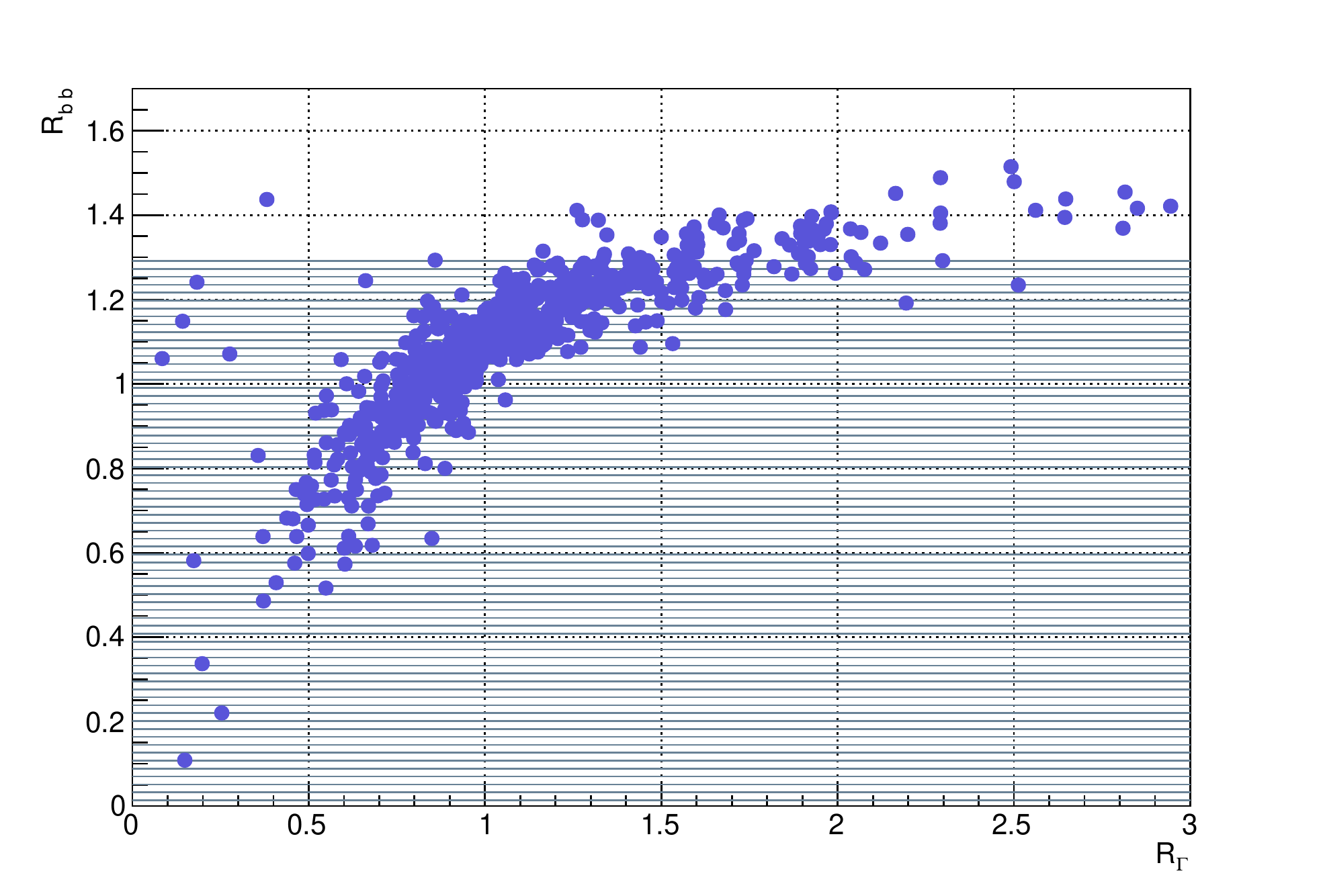}
\caption{Contribution to $R_{bb}$ as a function of $R_\Gamma$.}
\label{fig:fig2}
\end{figure}

\begin{figure}[ht!]
\includegraphics[scale=0.8]{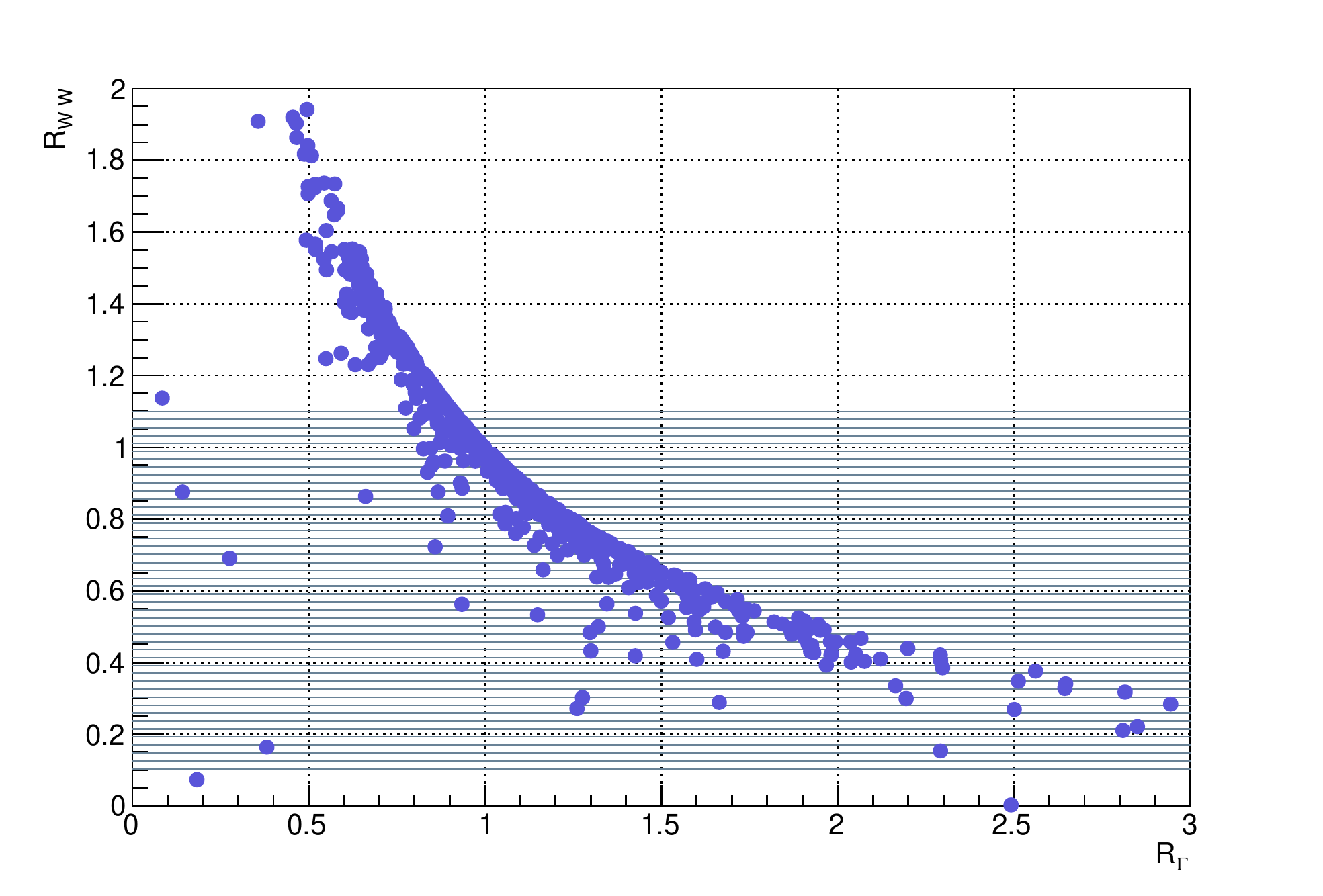}
\caption{Contribution to $R_{ww}$ as a function of $R_\Gamma$.}
\label{fig:fig3}
\end{figure}

In  Figure~\ref{fig:fig1} we present the values obtained for $R_{\gamma \gamma}$ as a function of the relative width of the light Higgs compared to the SM one. Different points correspond to different sets of the parameters in the potential that satisfy all the above conditions. The shaded region with horizontal gray lines corresponds to the $1- \sigma$ region of the best value in~\eqref{Rbestvalue}. The rest of the scalars in the model have masses either around of the light charged scalars, one CP-even and one CP-odd, or more massive by hundreds of GeV, the other charged, CP-even and CP-odd scalar. Points where the width of the Higgs is smaller than the one of the SM correspond to cases where the decay to bottoms is suppressed as shown in Figure~\ref{fig:fig2}, there are  also points where the width is close to one but they have reduced coupling to the bottom, those correspond to cases where the contributions from the charged higgses enhance the diphoton rate. It can be noted that the behavior of the diphoton rate and the bottom rate is opposite. For completeness we also include to rate of decay into gauge bosons in Figure~\ref{fig:fig3}. It can be seen that the region where the three rates agree with the data is when the lightest higgs has a relative width between 0.7 and 1.

As mentioned above, the points in Figure~\ref{fig:fig1} satisfy all the conditions discussed earlier, in particular the bounds from $K-\overline{K}$ mixing. However, having light charged Higgses can contribute significantly to several flavor changing processes not present (or extremely suppressed) in the SM such as $b \to s \gamma$: according to the calculation of the radiative decay $B\to X_s \gamma $  at Next-to-Next-to-Leading Order (NNLO) in the SM, leading to the prediction 
BR$(B\to X_s \gamma)_{\rm SM}= (3.15\pm 0.23) \times 10^{-4}$ \cite{Misiak:2006zs}, and  the current average of the measurements by CLEO  \cite{Chen:2001fja}, Belle \cite{Abe:2001hk,Limosani:2009qg}, and BaBar \cite{Lees:2012ym,Lees:2012wg, Aubert:2007my} reads 
BR$(\bar{B} \to X_ \gamma)|_{E_\gamma > 1.6 \,\, {\rm GeV}} = (3.37\pm 0.23) \times 10^{-4}$. 
 
Using this information and some previous studies we check if the solutions we have found so far can satisfy the constraint of the process $b \to s \gamma$: first we use the study of the so-called "democratic 3HDM"~\cite{Grossman:1994jb}, then we take the analysis of models with similar Higgs-fermion couplings, either with natural flavor conservation (NFC)~\cite{Akeroyd:1994ga,Akeroyd:2012yg,Aoki:2009ha, Pich:2009sp} or with FCNC induced by four-zero Yukawa textures~\cite{HernandezSanchez:2012eg}. In all these cases, and in our model, it is possible to obtain the generic interaction Lagrangian for both charged Higgs bosons  ($H^+_i$ , $i=$ 1,2) with the fermions:
\begin{eqnarray}
{\cal L}^{\bar{f}_u f_d H_i^+}  =
-\bigg\{\frac{ \sqrt{2} }{v} \overline{u}
\bigg(m_{d} X_{i} {P}_R +m_{u} Y_{i} {P}_L \bigg)d \,H^+_i
+\frac{\sqrt{2} m_{l} }{v} Z_{i} \overline{\nu_L} {l}_R  H^+_i  + H.c. \bigg\} .
\label{lagrangian-f}
\end{eqnarray}
where the parameters  $X_{i}$, $Y_{i}$, $Z_{i}$ are given by the Yukawa texture studied in our model. When the $b \to s \gamma$ process is considered we obtain the following bounds:
 \begin{eqnarray}
 |Y_1|^2+ |Y_2|^2 < 0.25, \,\,\,\,\,\,\,   -1.7< Re[X_1Y_1^* +X_2 Y_2^*] <0.7
 \end{eqnarray}
Both bounds apply when $m_{H_1^+} \sim  m_{H_2^+} $ and both charged Higgs bosons are light ($150$~GeV $\leqslant m_{H_i^+} \leqslant 350$~GeV). When there is a light charged Higgs and a  heavier one, the terms $X_i$ and $Y_i$ are absent. We have applied these bounds to our analysis and find that there are solutions compatible with them. A complete and comprehensive analysis for this and other related models is currently under preparation~\cite{3hdm}.

In this letter we considered the recent diphoton rate excess presented by ATLAS and CMS in the context of a renormalizable flavor model with three Higgs doublets and based on $Z_5$. We found that the model can accommodate the observed excess as well as all flavor observables including constraints from FCNC and $b \to s \gamma$.

\acknowledgments
This work was supported in part by PROMEP and CONACYT. AD was partly supported by the National Science Foundation under grant PHY-1215979.

\end{document}